# Playable Pressure:

# Affective Dramaturgy and Selective Realism in the Design of a VR Emergency-Response Serious Game

Jan K. Argasiński

*Institute of Applied Computer Science,*

*Faculty of Physics, Astronomy and Applied Computer Science,*

*Jagiellonian University in Krakow, Poland*

*& Nano Games sp. z o.o., Krakow, Poland*

*jan.argasinski@uj.edu.pl*

## Abstract

*Professional simulations stage not only procedures but models of what should command attention, which emotions belong in competent practice, and whose distress becomes part of the task. This article develops affective dramaturgy through critical design-document analysis of a virtual-reality emergency-response project. The corpus comprises two non-public production records. We identify six families of specified pressure and examine how sensory staging, proximity, trigger authority, task conflict, and response allocation imply a selectively receptive triage professional. The documented design can make emergency work morally and socially crowded, yet it can also turn grief, vulnerability, and mental-health-coded behavior into adjustable difficulty. We propose answerability at two levels—in-play response and post-play debriefability—and derive case-based questions about occupational purpose, representation, adaptation, and accountability. These questions are sensitizing propositions, not a validated framework for user effects or emergency practice.*

**Serious Games Stage Feeling as Well as Action**

A rescuer crosses an airport apron, checks breathing and circulation, and ties a colored triage band to a casualty. The procedure is orderly. The specified surrounding scene is not. A commander watches from close range and criticizes the work. A bystander records on a phone. Elsewhere, somebody screams, a relative presses for help, and a distant explosion demands attention without changing the documented clinical task. The records define these events as reusable patterns. Some are linked to player action or elapsed time. Others could be introduced by a researcher or instructor, or by a proposed physiological feedback loop.

The records model more than triage. They offer a version of what pressure should look like and how a professional should carry it. The implied rescuer must distinguish claims that matter now from those that can wait or should be resisted. These are cultural judgments about competence, authority, care, and composure, even when they appear in the practical language of scenario design.

This approach follows a path already established: Pelletier and Kneebone (2016) argue that medical simulation is a cultural practice with its own aesthetics, narrative conventions, and politics of professional labor. Their account shifts attention away from outcomes alone and toward the world a simulation asks participants to sustain. Sparrow et al. (2018) make a related point about professional training: simulations matter partly because they address dispositions and values, not only narrow physical skills. The present case brings those questions to the documented design of a VR emergency game whose pressure events are modular and whose bioadaptive layer remained a proposal in the analyzed corpus.

The scholarship also offers a way to study affect without pretending to observe an inner state. Dormann et al. (2013) show how affective learning can be organized through recurring game-design patterns, while Frome (2019) distinguishes emotion arising from interactivity in general from emotion organized by gameplay. Those distinctions matter here. A headset and embodied interface do not make a scene emotionally meaningful by themselves. Meaning develops through timing, proximity, task conflict, representation, and the actions the game makes available.

The case comes from a Polish industry R&D project aimed at professional emergency-service training, especially for firefighters and paramedics. Its non-public records describe two airport accident settings, ten planned scenario variants, forty pressure patterns, and twenty patterns given simplified minimum requirements because they were expected to prompt readily observable reactions. That priority rule is part of the evidence rather than a neutral sampling frame. The records cannot tell us what a player felt or learned, whether every specification was implemented, or how an autonomous controller performed. Project-archive development stills in Figures 1–5 illustrate visible staging only; they are not treated as evidence of speech, sound, triggers, interaction, or user response.

*We use affective dramaturgy as a bounded, middle-range concept for a documented relation among a represented source of pressure, its sensory and spatial staging, the timing and authority that activate it, its alignment or conflict with the player's task, and the response path the design provides or withholds. Here, playable does not mean solvable. Pressure becomes part of play when its appearance, proximity, task relation, and response are organized through rules, even when the intended action is justified deferral or resistance. The question is not whether the project successfully produced stress, but what professional subject its documented arrangement brings into view.*

The analysis finds a recurring tension in the records. The design can render emergency work morally and socially crowded rather than merely technical. Yet its most vivid pressures often arrive through distressed people whose claims are easier to trigger than to answer within the specified interaction model. Affective dramaturgy helps describe both tendencies at once: an enlarged field of professional attention and a narrowed set of relations through which some claims can be addressed.

## From Affective Patterns to Affective Dramaturgy

### Patterns as a Design Language

Game-design patterns describe recurring relations in play rather than isolated assets (Björk & Holopainen, 2005). In serious games, they help a team discuss how an event works across more than one scene. Dormann and colleagues brought this approach into the affective domain by identifying patterns that support emotional representation and socioemotional interaction (Dormann et al., 2013). Earlier work on affective patterns in serious games connected the same design language to affective computing and evidence-centered design (Argasiński & Węgrzyn, 2019).

Reusability is useful in production, but it has cultural consequences. Consider a frightened victim who can be placed beside the player in several scenarios. Turning fear into a pattern requires decisions about voice, animation, distance, duration, and response. It also requires a decision about who will embody that fear. A production tool thus becomes a small model of social life. It decides what kind of distress is legible, how insistently it reaches the player, and whether it remains part of a person or travels as a detachable difficulty setting.

This is where the present article departs from an instrumental account of patterns. The question is not simply whether an event is reusable or likely to sustain engagement. It is what reusability does to the represented person, the player's professional role, and the game world's account of emergency work. De Smale and colleagues make production choices central to moral game design, showing that ethics cannot be separated from the contexts in which gameplay is assembled (de Smale et al., 2019). The production documentation examined here offers a narrower view than interviews or ethnography, but it still reveals how a proposed system joined representation to rules.

### Affective Dramaturgy

Affective dramaturgy begins with composition. Its unit is not an isolated asset or pattern but a pressure event positioned within a task sequence. A scream from an unassessed casualty has one specified relation to triage; the same scream placed beside a rescuer already treating someone else has another. Criticism from a uniformed commander differs from a bystander complaint because the represented voices carry different authority. An event made eligible by a physiological estimate also differs from one chosen openly by an instructor, even if its animation were identical. ‘Dramaturgy’ names this composition across time and relation, not a theatrical narrative or an inference about players’ feelings.

Pattern analysis identifies reusable design relations (Björk & Holopainen, 2005; Dormann et al., 2013; Argasiński & Węgrzyn, 2019), while formal and procedural analysis identifies rules, affordances, and repeated action (Consalvo & Dutton, 2006; Fernández-Vara, 2019; Matheson, 2015). Affect scholarship treats feeling as distributed across bodies, interfaces, spaces, and relations (Ahmed, 2004; Anable, 2018; Shaw & Warf, 2009), and cautions against attributing emotion to interactivity alone (Frome, 2019). Affective dramaturgy contributes the narrower comparative question of how staging, trigger authority, task conflict, and response allocation combine to make one demand a professional relation, another an environmental constraint, and another a spectacle to resist. We use it as both an artifact property and an analytic lens, not as evidence of designer intention, production negotiation, or experienced emotion.

*We introduce answerability as an operational term with two temporal levels. In-play answerability asks whether the player can acknowledge, act on, refuse, delegate, or record a demand within the scenario. Post-play debriefability asks whether learner and instructor can reconstruct why the event appeared, what choices existed, and what the design excluded. Debriefing can make an in-play restriction discussable; it does not retroactively create an affordance. Limited response is not automatically a defect: in triage, intentional nonresponse may be role-appropriate when tied to a specified protocol or learning objective. Answerability is not a synonym for agency or affordance; it evaluates whether a particular demand is paired with a professionally intelligible response or justified deferral.*

Figure 6 maps this case-derived analytic sequence. A represented source of pressure is specified through sound, movement, visibility, and distance; timing and trigger authority position it in relation to the task; and available or withheld responses imply what should be noticed, managed, resisted, or prioritized. A separate debrief may reopen that implied feeling rule for discussion, but it remains analytically distinct from action available during the scenario.

## Feeling Rules and the Implied Professional

Hochschild's account of emotional labor describes how occupations organize appropriate feeling and display (Hochschild, 1983). Those rules are learned socially, as studies of emotion management in fire stations make especially clear (Scott & Myers, 2005; Wharton, 2009). A serious game can participate in that education even when it never names emotion as a learning objective. It does so by rewarding certain forms of attention and making others difficult or impossible.

The records do not document the actual emotional culture of emergency services, and this article makes no such claim. They stage a designer-produced version of competent conduct in a Polish industry R&D setting. Games have long been analyzed through the abilities and dispositions they presume in players (Aarseth, 2003; Juul, 2005); a professional simulation adds an occupational role. Here, competence is implied through triage actions performed amid scrutiny, interruption, and moral address. 'Professional culture' therefore refers to the model encoded in these records, not to Polish emergency-services culture generally.

The implied professional is neither detached nor openly overwhelmed. The rescuer should notice relevant suffering without allowing every vivid event to reorganize the task. This controlled receptivity is one of the records' strongest cultural propositions. It frames composure as a practical achievement, yet may also individualize resilience because communication, mutual support, and institutional resources are less fully specified than endurance.

**Selective Realism and System Ethics**

Realism in games comes from selection. Pötzsch describes selective realism as the filtering through which games make some parts of violence palpable and push others beyond the playable frame (Pötzsch, 2017). The same logic applies to professional simulation. A game can render equipment and bodily states in detail while saying little about teamwork, organizational failure, recovery, or the long aftermath of an incident. Fidelity is always fidelity to something, chosen for a purpose (Hontvedt & Øvergård, 2020).

Pelletier and Kneebone show why this matters in medical simulation: a simulated event does not merely reproduce practice but rearranges how practice can be understood (Pelletier & Kneebone, 2016). Weines and Borit likewise ask analysts to consider a game world's accuracy, comprehensiveness, balance, and internal consistency rather than treat correspondence with reality as a single scale (Weines & Borit, 2022). For the present case, the key issue is balance. The specification foregrounds a crowded field of voices and bodies, visually illustrated by Figure 3, while organizational preparation and shared responsibility are comparatively thin in the focal records.

Ethical analysis has to reach beyond content. Donald's study of war-game design connects representational choice to the responsibilities a system makes actionable (Donald, 2019). De Smale and colleagues place moral design within production, where technical and organizational choices shape the resulting experience (de Smale et al., 2019). Montefiore and Formosa sharpen the distinction by separating objectionable content from objectionable systems (Montefiore & Formosa, 2026). A depiction of grief is not responsible simply because grief belongs in emergency work. Its meaning depends on why it appears, how often it can recur, what the player can do, and what happens afterward.

## Case, Corpus, and Critical Design-Document Analysis

### Case and Corpus

The case comes from a Polish industry R&D project aimed at professional emergency-service training, especially for firefighters and paramedics. The scenario positions a first-person member of a rescue team at an airport mass-casualty incident and centers correct triage classification and casualty assistance. Planned settings involve an airport bus colliding with an aircraft or being overturned by jet blast. The records do not establish one commissioning institution, trainee qualification level, national protocol, incident-command role, or deployment workflow. Communication, delegation, documentation, and de-escalation discussed below are therefore analytical design possibilities whose appropriateness would depend on role, protocol, team configuration, and scenario phase.

The specified VR architecture uses first-person head-tracked viewing in an HTC Vive, room-scale walking within a marked safe area, and controller-ray teleportation. A controller ray selects characters and objects, while a touchpad menu exposes distance-dependent actions; close range enables clinical checks, history taking, triage-band assignment, cardiopulmonary resuscitation, and object use. These constraints make proximity and head-directed attention mechanically consequential in the documented design. The records do not report field of view, refresh rate, spatial-audio implementation, avatar embodiment, accessibility settings, or cybersickness controls, and no hardware configuration was tested in this study.

The corpus comprises two non-public Polish-language records. The primary source is a 34-page scenario-and-interaction specification; it documents the player role, movement and interaction model, two accident settings, ten planned variants, character states, trigger rules, a forty-pattern inventory, and simplified requirements for twenty prioritized patterns. The second is a seven-page system overview describing the proposed bioadaptive architecture. The overview contextualizes but does not independently validate the specification. These are the complete records analyzed here.

## Critical Design-Document Analysis

This is critical design-document analysis, grounded in document analysis and informed by formal and procedural game analysis (Bowen, 2009; Consalvo & Dutton, 2006; Fernández-Vara, 2019). ‘Specified’ and ‘planned’ refer to document content; ‘proposed’ refers to the physiological loop; and ‘shown in a development still’ refers only to visible spatial or visual features. Formal analysis is limited to rules, menus, spatial instructions, and trigger relations described in the records; procedural analysis reconstructs how those elements would relate if instantiated. The production-oriented contribution is therefore bounded to documented design choices, not negotiation, labor, private intention, technical performance, or experienced play (de Smale et al., 2019).

The unit of inventory was a documented pattern specification; ‘event’ denotes a possible instantiation. We coded all forty for selection status, carrier or source, modality, spatial rule, trigger, task relation, and available response, then applied deeper interpretive codes to the twenty with fuller specifications: implied professional demand, implied feeling rule, selective-realism function, answerability, and representational risk.

The priority split is part of the evidence, not a neutral sampling frame. The records state that twenty patterns received fuller minimum requirements because they were expected to prompt readily observable reactions. We therefore do not present the subset’s sensory vividness as an independently discovered property; we ask what cultural distribution follows when observability becomes a production criterion. The broader inventory also contains time pressure, casualty load, flawed briefing, engine noise, fuel leakage, and threat to the rescuer. Their thinner specification is consistent with salience-oriented selection, but feasibility, cost, sequencing, reuse, and evaluation convenience are alternative explanations the records cannot adjudicate. We call these pressures less developed at this stage, not intentionally excluded.

Table 1 presents six interpretive families for the twenty prioritized pattern specifications. Each pattern receives one primary family for readability, although operations can cross families; the taxonomy is exhaustive only for this prioritized set and is not a psychological scale or designer-authored classification. Stable identifiers, the derived forty-pattern inventory, coding guide, and reflexive scope decisions appear in the Supplemental. Auditability rests on those IDs, logged decisions, separation of description and interpretation, and negative cases rather than a reliability coefficient.

The authors had insider access to the focal project, and some participated in project work. The access enabled approach and vocabulary familiarity but also risks selective emphasis, defensive interpretation, and retrospective coherence. We therefore rely on records rather than recollection, distinguish specification from interpretation, log coding decisions, test claims with negative cases, and make only corpus-relative absence claims.

### A Grammar of Playable Pressure

The twenty prioritized specifications share a practical structure. Each gives pressure a source, sensory form, location, trigger, and relation to the triage task. Table 1 groups them by primary family and supplies stable IDs; Figure 6 shows operations that cut across the groups. Several events could be attached to any suitable casualty or bystander, and many are specified near the user. Reassignability and proximity make pressure portable across planned scenes.

### Social Scrutiny and Institutional Surveillance

Five specifications make the rescuer's performance public. A commander watches from nearby and criticizes (Figure 2). Bystanders comment, hurry the player, ask intrusive questions, or follow with a phone. A media crew arrives with cameras and microphones. As documented, these events do more than add noise: they attach professional judgment to an audience.

The audience is not uniform. A commander represents hierarchy, while a reporter or phone-bearing witness brings reputational exposure. Yet the specified menu gives these figures similar practical weight. It supports detailed clinical checks and a general request that people move away, but documents no equally developed path for briefing media, explaining triage, challenging interference, or delegating communication. Those actions may not belong to every trainee role; their absence therefore identifies a representational boundary to discuss, not proof of a protocol defect. Scrutiny is carefully specified, while professional communication is comparatively thin.

This documented arrangement has a double edge. Public observation can make accountability part of a scenario, which is relevant to emergency work. But criticism paired only with persistence frames the social world primarily as a concentration test. If implemented as specified, the rescuer would remain visible and composed while having few ways to shape the encounter.

**Victim Distress and Relational Claims**

Five specifications make suffering audible and relational. Casualties scream or cry; relatives approach (Figure 3); an early-pregnancy appeal requests priority; and a casualty seeks help for a missing child while holding a toy. Voice, kinship, and bodily vulnerability are positioned against the abstraction of triage. Figure 4 illustrates the visible staging associated with the missing-child specification.

These details can humanize a procedure built around categories and colored bands by specifying attachments and fears around a triage decision. At the same time, some identities and relationships can travel between otherwise interchangeable characters. A nearby figure can be assigned the role of relative; a conscious woman can carry the pregnancy event; a toy can mark a casualty as a parent searching for a child. The claim becomes vivid before the person becomes specific.

Again, the documented menu matters. It specifies casualty examination but no equally developed route for recording the missing child, explaining priority to a relative, or acknowledging the pregnancy concern in a tailored way. Whether those actions would be role-appropriate depends on protocol and team configuration. Analytically, the asymmetry shows that moral address is more fully staged than its possible professional handling; it should not be treated simply as an interface error.

## Clinical Deterioration and Moral Urgency

Hemorrhage and sudden death are specified as changes to a casualty rather than surrounding interruption. A pool of blood appears beneath one body; another loses signs of life and changes position. If instantiated with task consequences, these events could destabilize an earlier assessment and make time and reassessment part of the professional problem. Figure 3 illustrates visible bodies and spacing.

The specification gives visual change more detail than clinical development. Blood appearance is explicit; a revised pathway of care is less so. Death is described as an abrupt state transition rather than a sequence of cues that might support intervention or uncertainty. By contrast, the wider inventory includes time pressure and deterioration after assessment, both of which would place more weight on longitudinal judgment. Given the observability-based priority rule, this contrast shows what that criterion made easier to specify, not an independently discovered preference or verified runtime behavior.

Conspicuous bodily change may be appropriate for a specified learning objective. The analytic concern is proportionality: where visible force is documented more fully than task consequence, the record supports a claim about design emphasis, not about what a player actually noticed or learned.

## Environmental and Perceptual Disruption

Storm and fog place pressure in the environment rather than in a vulnerable person. The records specify rain building into thunder and lightning, and fog narrowing visibility. If instantiated, these conditions could affect seeing, hearing, navigation, and reassessment, but the corpus does not establish their runtime fidelity or user effects.

Their occupational relevance depends on the specified relation to work. Reduced visibility becomes more than atmosphere if it changes casualty search or reassessment; weather remains primarily scenic if the clinical task proceeds unchanged. The records develop audiovisual change more fully than downstream effects, but the patterns identify another route for affective design: uncertainty can be located in incomplete information, spatial conditions, and task dependencies.

The broader inventory points in the same direction. It includes engine noise, fuel leakage, flawed briefing, and danger to the rescuer. Those candidates locate pressure in the ecology of emergency work and show that the documented design repertoire was not limited to distressed bodies, even though these alternatives received thinner specification at this stage.

**Exceptional Spectacle and Improbable Interruption**

Two specifications describe conspicuous, unlikely events. A large explosion occurs far enough away that the rescuer need not leave the current site. In another, a giraffe walks calmly into the accident area while people react around it. The records label the latter a 'black swan' pattern; we treat that phrase only as the project's label, not as a theoretical category. Figure 5 illustrates the visible giraffe scene. The two events differ in tone, but both separate attentional capture from a change in the documented triage task.

Surprise belongs in emergency work: incomplete information and unforeseen hazards can require plan revision. Here, however, the explosion is specified to be noticed without changing the local task. If implemented as specified, the event would ask the player to resist a sensory claim rather than respond to a new danger. The giraffe makes that operation comic and explicit. These records thus permit novelty to stand in for uncertainty, although they do not show how users interpreted it.

The question is not whether an improbable event is realistic. Selective realism always includes invention. The more precise question is what occupational purpose the invention serves. Resisting irrelevant distraction can itself be a legitimate objective; alternatively, surprise may change judgment, relation, or responsibility. Without a stated objective or task consequence, the documented event remains ambiguous rather than automatically illegitimate.

**Behavioral and Interactional Disruption**

Four specifications place pressure in difficult interaction: aggression by a casualty, aggression by a bystander, panic, and a pattern the records frame in psychiatric terms. The last combines hallucination-like behavior, disordered speech, and erratic movement. We paraphrase rather than reproduce the source label and refer to this as mental-health-coded behavior; the phrase identifies the design’s framing and does not diagnose the character. Any suitable casualty could carry some of these behaviors.

Such encounters could support risk assessment, de-escalation, psychological first aid, or an adapted clinical examination, depending on trainee role and protocol. The corpus specifies threatening speech and disruptive movement more fully than those practices. This bounded formal imbalance can position mental-health-coded difference primarily as attentional pressure; it cannot support claims about mental illness, actual patients, or player attitudes.

Changing a label alone would not resolve the representational issue. The relation among described animation, reassignability, proximity, and differentiated action matters. The records make threatening or erratic behavior portable while specifying few tailored responses beyond persistence. This creates a procedural risk of reducing a represented person to disruption; whether that risk materialized in play is outside the evidence.

Across the six families, content does not act alone. A relative, commander, storm, or giraffe becomes pressure through documented timing, proximity, task competition, trigger authority, and response allocation. Affective dramaturgy names that composition.

**The Implied Professional**

The scenario's explicit work is triage: the specified player approaches casualties, checks signs, gathers information, classifies urgency, and provides limited aid. Its implicit work is to protect that sequence against interruption. The records do not assign a formal score for composure, but their continuing action demands imply composure while the surrounding scene becomes louder, more public, or morally complicated.

This makes attention both practical and ethical in the implied role. A scream, phone, and relative's plea are all specified as salient, yet triage cannot give every salient event equal priority. Separating urgency from vividness is a meaningful professional problem. The model becomes narrower where different people receive few distinct response routes and are positioned as one broad category of interruption.

The implied rescuer is expected to remain receptive without losing order. 'Staying calm' is too simple a description: the role should notice clinical change and human need, resist task-irrelevant spectacle, and translate claims into ranked action. The documented design frames this controlled receptivity as skilled work, while also positioning professional feeling as an individual resource repeatedly exposed to pressure.

Several authorities are specified around the player: the commander brings hierarchy, media bring visibility, casualties and relatives bring moral claims, and a researcher or instructor can control event timing. The proposed bioadaptive layer adds another authority. The overview sketches a chain from BITalino-based cardiac and electrodermal acquisition, through an individual baseline and a scalar engagement estimate, to event eligibility when the estimate falls.

A scalar estimate should not be mistaken for direct access to emotion, and automation does not erase authorship. If an instructor introduces a distressed relative, trigger authority is visible; if a metric initiates the same event, escalation may appear to result from the learner's own insufficient engagement. A represented person's suffering then becomes a proposed actuator in a control loop. This interpretation concerns documented authority and representation, not proof that autonomous triggering occurred.

The implied cultural model is subtle. The records do not simply reward emotional suppression; they specify managed openness—sufficient sensitivity to notice what matters and sufficient control to preserve procedure. That proposition has occupational plausibility because emergency competence includes action amid uncertainty and emotion. In this corpus it remains selective, since team support, recovery, and institutional critique are comparatively weakly specified.

**Selective Realism and Modular Distress**

The prioritized specifications foreground public scrutiny, vulnerable voices, bodily deterioration, bad visibility, aggression, and panic. The wider inventory also acknowledges time pressure, casualty load, unreliable briefing, engine noise, fuel leakage, and danger to the rescuer. The difference lies in which patterns received fuller minimum requirements at this design stage, not in a verified implementation boundary.

Because anticipated observability helped define the prioritized set, its immediate sensory presence is partly built into selection. The analytical question is what cultural distribution that criterion operationalizes. Conspicuous people receive fuller specifications than coordination, resources, or uncertain information. Feasibility, cost, development sequence, reuse, and evaluation convenience remain alternative explanations; the records cannot establish private intention or a single cause.

The scenario centers a single first-person rescuer. Team coordination, mutual support, delegation, fatigue, recovery, and institutional aftermath receive comparatively little specification in the focal corpus. This corpus-relative imbalance makes endurance look personal: pressure is carried by the implied individual, while organization appears mainly through command and evaluation. It does not establish their absence from the wider project or from real emergency practice.

Modularity brings real benefits. Reusable events can support comparison across sessions, instructor control, and focused discussion among designers and domain experts. The concern is what gets lost when an event travels. Pregnancy, grief, mental-health-coded behavior, and a missing child can become detachable sources of intensity. Repetition may support rehearsal, but the documented relation can also script vulnerability as a familiar obstacle. The corpus does not establish what players learned from repetition.

Representational risk is not shared evenly. Fog carries no stigma, while mental-health-coded behavior may activate stigmatizing associations. Pregnancy and kinship may be professionally relevant, but can also operate as familiar shortcuts to urgency and grievability (Butler, 2009). Aggression can be a legitimate safety concern while still being specified in ways that hide causes and flatten possible responses. Ethical analysis therefore asks who bears pressure and what kinds of relation the documented action model permits.

In-play answerability provides one practical test. The specified interface is comparatively rich for examining a body and assigning a band, but sparse when a family member seeks information, a reporter demands access, or a distressed person might require de-escalation. Role-appropriate refusal may be the intended action, yet that justification is not made explicit. Post-play debriefability is a separate test: discussion can reopen an encounter only if its timing, trigger authority, available options, and excluded choices can be reconstructed (Crookall, 2010; Fanning & Gaba, 2007).

A more precise test of modular distress relates occupational purpose, representation, trigger control, in-play response, and reflective aftermath. Sudden death tied to reassessment could create a difficult clinical problem; the same event selected only to raise a scalar estimate would treat a body as a difficulty modifier. A relative could make triage ethically concrete or merely noisier. These are analytical contrasts, not claims about observed sessions. Selective realism becomes more accountable when such distinctions are documented and open to challenge.

### Case-Derived Questions for Responsible Affective Dramaturgy

This single documentary case does not yield an empirically validated or universal framework. It brings four tensions into focus—salience and task consequence, representation and response, trigger authority and accountability, and intensity and aftermath. Table 2 gathers eight sensitizing questions. Some follow directly from pattern contrasts; others, especially safety controls and governance of adaptation, are normative extensions prompted by what the records leave unspecified.

### Salience and Occupational Relevance

Salience and occupational relevance are not opposites. Resisting task-irrelevant distraction may itself be a legitimate training objective, so an event need not alter a casualty's state to be relevant. The design question is whether that objective and the expected response are specified. The distant explosion captures attention while leaving the local task intact; fog could matter differently if it changes search or reassessment. Purpose should therefore be stated before intensity is treated as evidence of value.

Affective specificity follows from the same point. Fear, grief, scrutiny, uncertainty, and surprise need not become interchangeable because a proposed controller uses one engagement estimate. An event library can preserve the difference among a person to acknowledge, a hazard to investigate, and a spectacle to resist, while recognizing that the appropriate action depends on role and protocol.

**Representation, Restricted Response, and Role**

The records give distressed characters strong presence but often a weak procedural future. Responsible representation requires more than careful wording or convincing animation. Depending on role and phase, a design might provide acknowledgment, explanation, boundary setting, help-seeking, documentation, or explicit justified deferral. These are normative possibilities rather than duties established by this corpus. Without either a response path or a protocol-based rationale for restriction, vulnerability remains visible mainly as pressure on the professional.

Answerability also changes representational risk. Mental-health-coded behavior becomes less reducible to disruption when a scenario specifies differentiated communication or care; a relative becomes more than noise when information exchange is recognized as part of the role. Conversely, explicit protocol-based nonresponse can also be meaningful. Representation gains specificity through the relation among action, justified restriction, and aftermath.

**Adaptation Does Not Outsource Responsibility**

Instructor control and physiological adaptation can both support responsive scenarios, but they govern when another person’s distress is introduced. That authority should remain traceable.

Event logs, bounded intensity, genuine pause and stop controls, cooldowns, instructor override, and a record of trigger authority would make escalation reconstructable. These are author-proposed requirements, not safeguards verified in the two records.

Interpretive humility and biometric governance are equally important. Physiological signals are uncertain indicators and should not be presented as transparent readings of feeling. The implementation documents should specify consent for adaptive sensing, data minimization or retention, access controls, trainee access to inferred-state records, differential-error review, distress or cybersickness stopping criteria, or post-session support. Designers and institutions remain responsible for the library, estimator, thresholds, timing, data practices, and representational choices.

**Intensity and Reflective Aftermath**

A difficult encounter can remain educationally and culturally thin if reflection ends with the headset session. A documented debrief could let learner and instructor revisit why a demand appeared, what the interface allowed, and which practices were excluded. It could also expose selective realism: what the scenario simplified, exaggerated, or left out.

Reflection should include the design's framing, not only learner performance. Asking why a relative could plead but not receive an explanation, or why an explosion demanded attention without changing the task, turns design choices into discussable objects. Debriefing cannot retroactively supply an in-play affordance or remedy harmful representation by itself; it can make the restriction and its rationale available for critique. The aim is to connect intensity to occupational purpose and accountable interpretation.

## Conclusion

The giraffe is the easiest specification to remember: it enters an airport-disaster scene, attracts represented attention, and changes nothing about the casualty in front of the implied player. Precisely because it is improbable, it makes the design logic visible. Pressure is assembled from a source, place, moment, trigger authority, task relation, and response path. The same analytic components organize the commander, crying casualty, phone camera, and sudden blood pool.

Reading those specifications together reveals a particular designer-produced model of professional competence. The records imply a rescuer who remains open to clinically and morally relevant information without letting spectacle, criticism, or distress dissolve procedural order. This is a serious but selective account of emergency work: individual composure receives more detail than collective support, and some vulnerable people receive more presence as interruptions than as participants in care. The claim is limited to the two-record Polish R&D case, not emergency culture generally.

Affective dramaturgy offers a way to examine documented choices without claiming access to player emotion. It directs attention to how a design stages pressure, distributes trigger authority, and pairs represented claims with response or justified deferral. As a single case, this analysis does not establish a transferable framework. It offers propositions for comparative research in health, safety, military, and workplace simulations, where the components may be refined, rejected, or extended. When pressure becomes programmable, a model of professionalism is written into the trigger.

**Funding**

This research was supported by the Polish National Centre for Research and Development under the grant 'Bioadaptable training simulator for critical infrastructure operators – research and preparation for implementation' (POIR.01.01.01-00-1131/17; the Smart Growth Operational Program, submeasure 1.1.1, received by Nano Games sp. z o.o.).

## Tables

### Table 1

*Six interpretive families of prioritized pressure patterns (n = 20).*

| **Family** | **Included prioritized patterns** | **Dominant operation and professional demand** | **Central cultural question** |
|---|---|---|---|
| Social scrutiny and institutional surveillance (n = 5) | P03 commander observation and criticism; P06 bystander comments; P08 phone recording; P09 media; P18 intrusive bystander | Makes performance public and tests work under observation, criticism, and reputational exposure. | Does competence include communication and boundary management, justified resistance to scrutiny, or both? |
| Victim distress and relational claims (n = 5) | P04 screaming; P05 crying; P12 victims' relatives; P19 early-pregnancy appeal; P20 missing child | Uses voice, kinship, and proximity to turn vulnerability into an immediate claim on attention. | When is vulnerability paired with a role-appropriate response or justified deferral, and when is it mainly an intensity source? |
| Clinical deterioration and moral urgency (n = 2) | P07 hemorrhage; P13 sudden death | Specifies a bodily-state change that could destabilize earlier triage decisions. | Does the event alter professional judgment, or is visual change developed more fully than task consequence? |
| Environmental and perceptual disruption (n = 2) | P10 rain and storm; P11 fog | Changes atmospheric and perceptual conditions without assigning pressure to a vulnerable person. | Is uncertainty linked to task consequences or specified mainly as scenic intensity? |
| Exceptional spectacle and improbable interruption (n = 2) | P14 distant explosion; P15 giraffe | Captures attention through salient but documented task-extraneous surprise. | Is resistance to distraction an explicit occupational objective, or does spectacle remain weakly tied to purpose? |
| Behavioral and interactional disruption (n = 4) | P01 aggressive casualty; P02 aggressive bystander; P16 mental-health-coded behavior; P17 panic | Specifies behavior as a pressure on composure while giving less detail to differentiated care or de-escalation. | Are represented people framed as patients needing care, interactional challenges, or obstacles, and what role-specific response is available? |

**Table 2**

*Case-derived and normative questions for reviewing affective dramaturgy.*

| Question | Design prompt |
|---|---|
| Occupational purpose | What judgment, communication practice, responsibility, or justified resistance does the event make available beyond generic salience? |
| Representational accountability | Who carries the pressure, what identity or vulnerability is mobilized, and does the design preserve specificity rather than use a person as shorthand? |
| Affective specificity | Does the event library distinguish grief, uncertainty, scrutiny, surprise, moral conflict, and hazard even if adaptation relies on a scalar estimate? |
| Proportionality and learner control | Are intensity, timing, repetition, pausing, stopping, escalation, and instructor or automated authority bounded and documented? |
| In-play answerability | Can the player acknowledge, act on, refuse, delegate, record, or explicitly defer the demand in a way appropriate to role and protocol? |
| Post-play debriefability | Can learner and instructor reconstruct the event, trigger authority, sequence, available choices, rationale, and design exclusions afterward? |
| Selective-realism disclosure | Does documentation identify what the simulation includes, simplifies, intensifies, and excludes in relation to its purpose? |
| Adaptation, interpretation, and governance | Are inferred states treated as uncertain, are consent and data safeguards specified, and who remains accountable for the estimator, thresholds, event policy, override, representation, safety, and aftermath? |

## Figures

**Figure 1:** *Illustrative project-archive development still of the airport scene.*

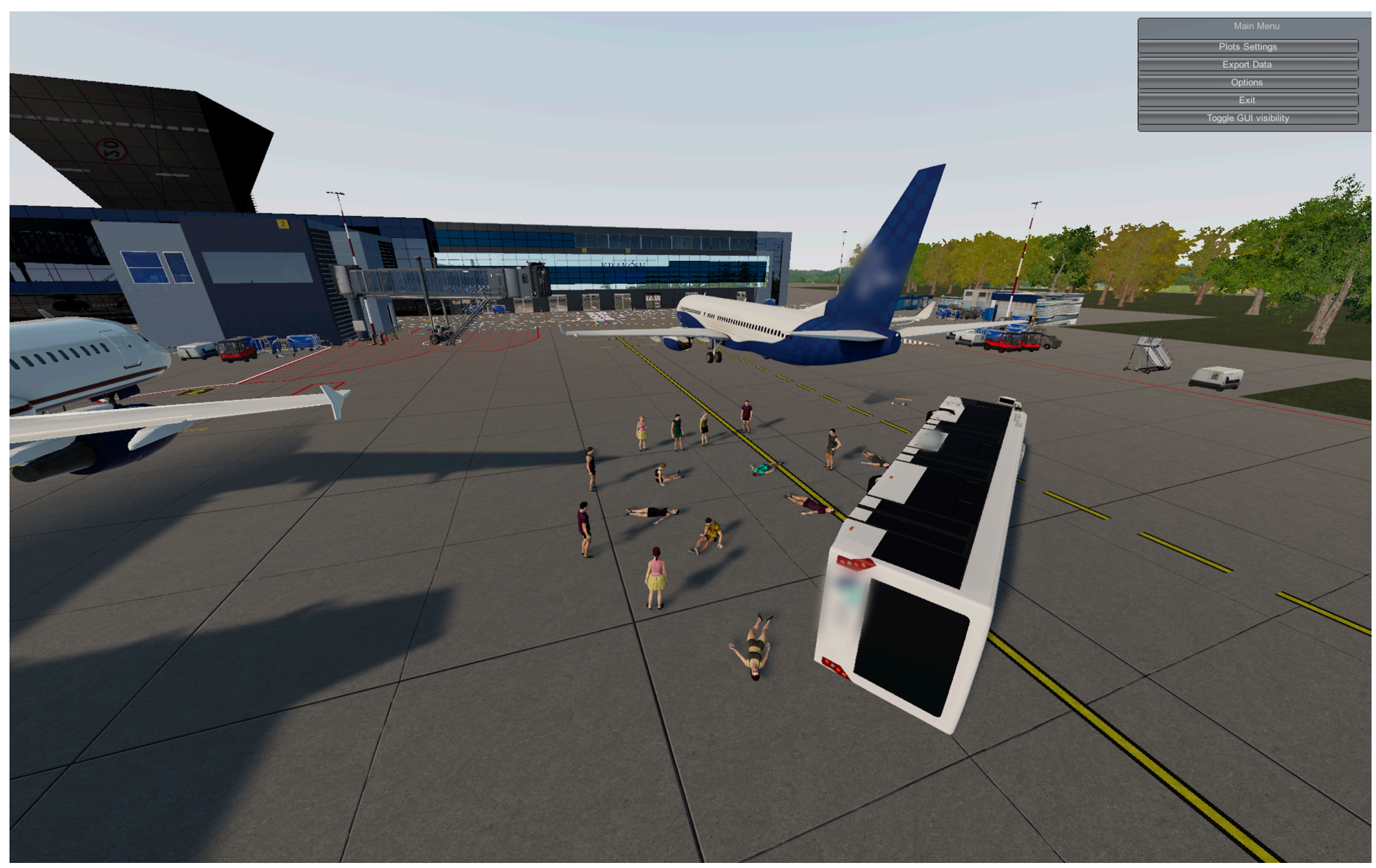

**Figure 2:** *Illustrative project-archive development still of the uniformed-commander event.*

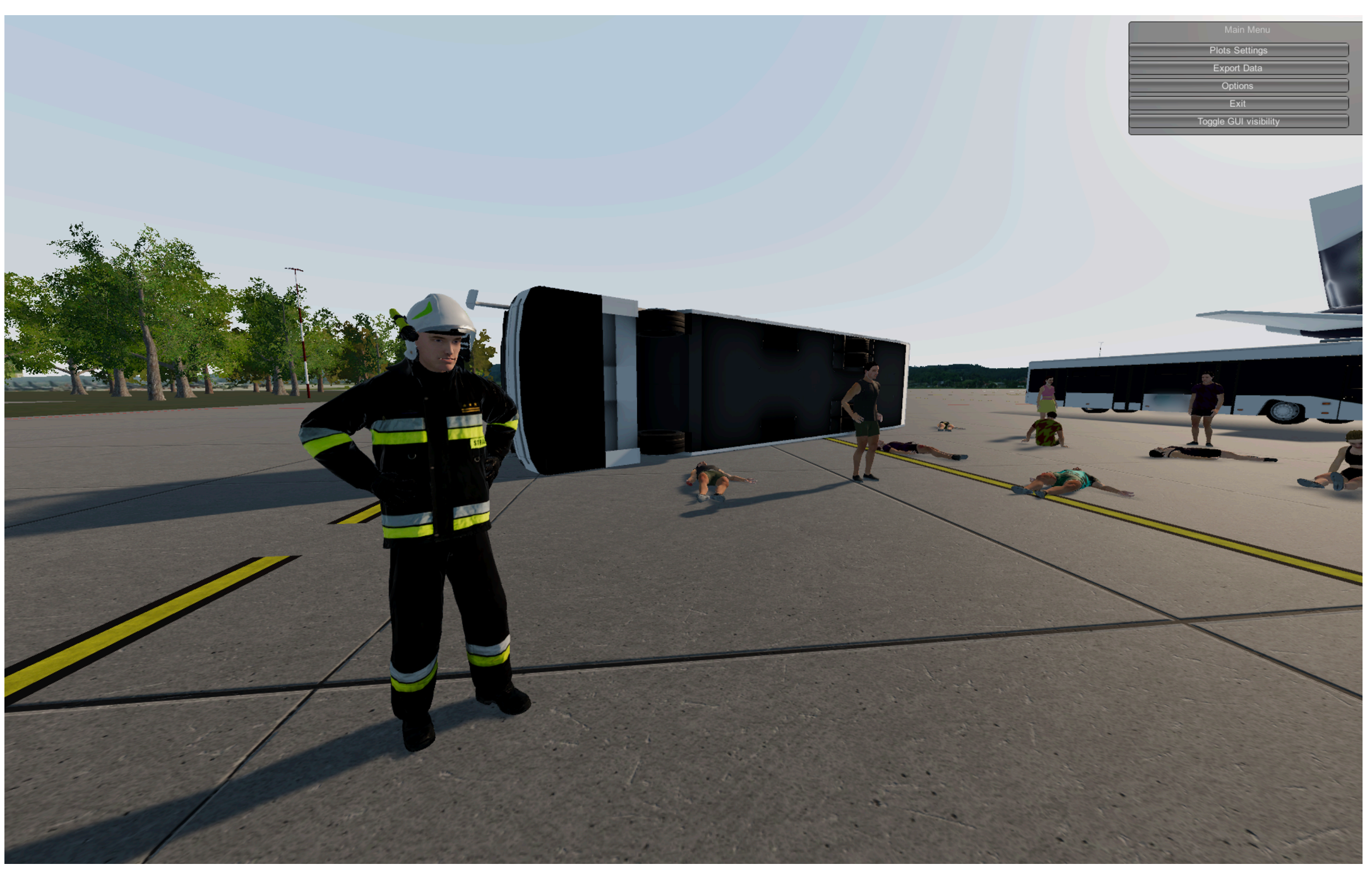

**Figure 3:** *Illustrative project-archive development still of casualties and bystanders.*

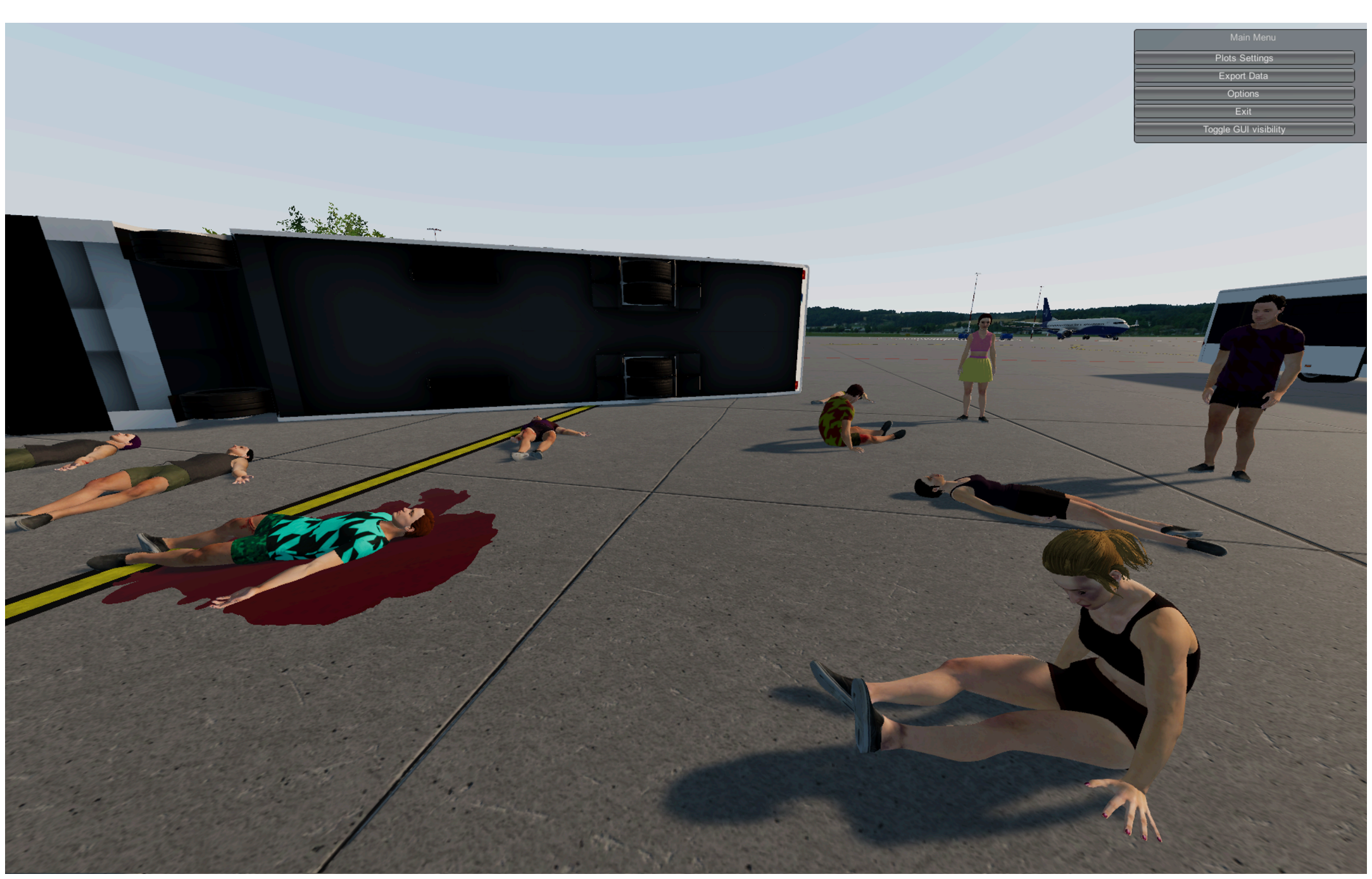

**Figure 4:** *Illustrative project-archive still associated with the missing-child specification.*

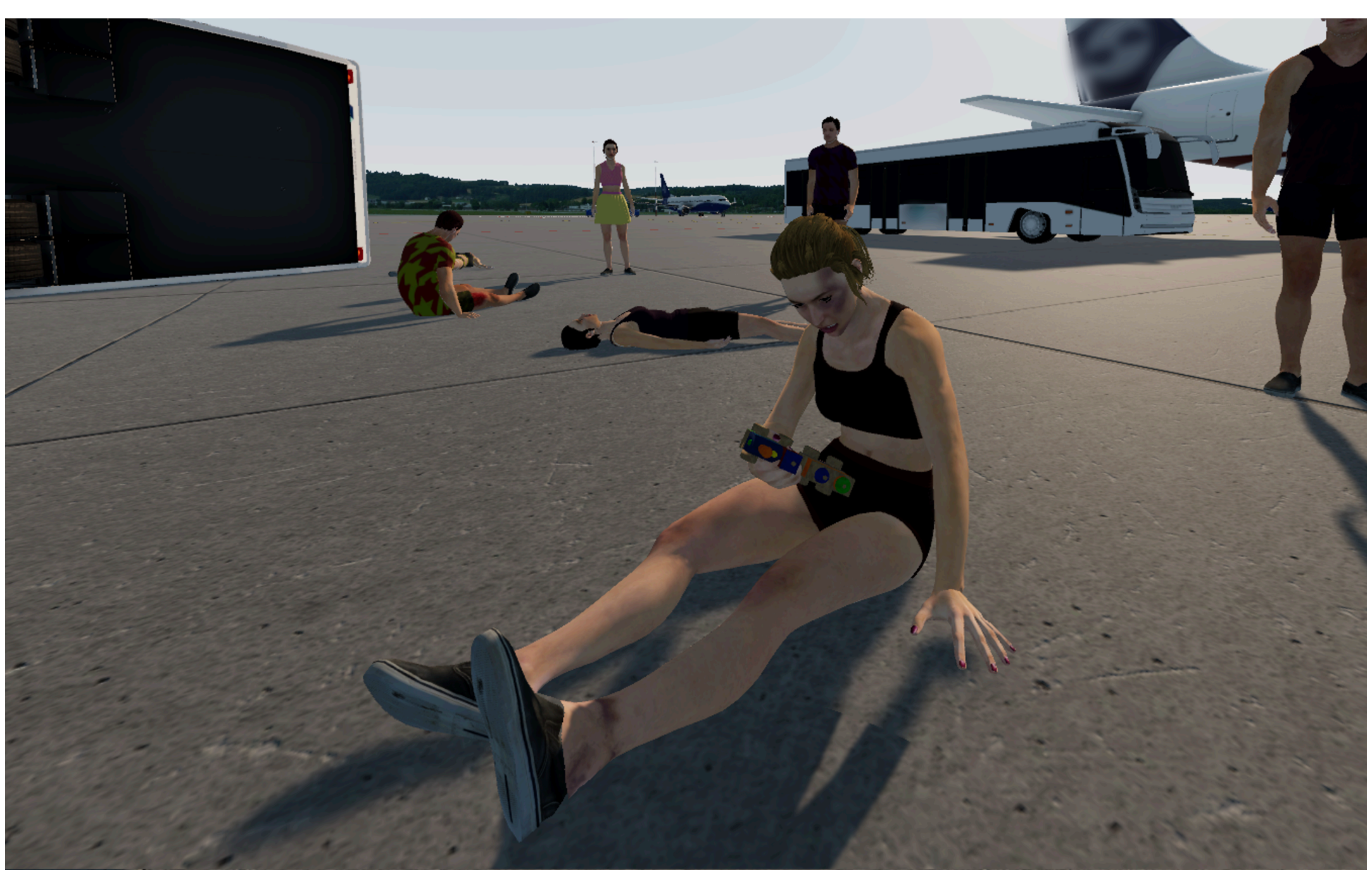

**Figure 5:** *Illustrative project-archive development still of the event labeled 'black swan'*

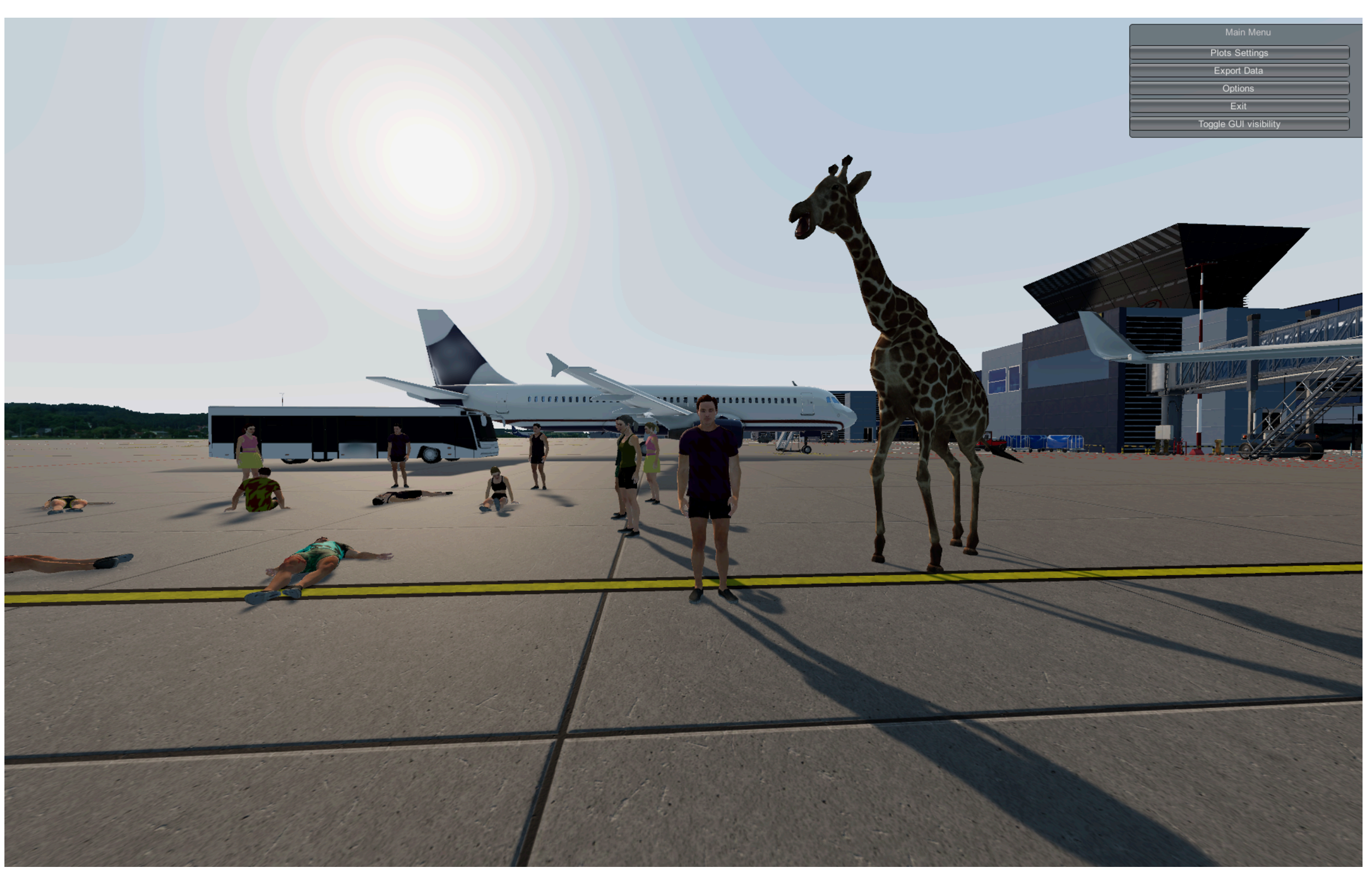

**Figure 6:** *Affective dramaturgy as a case-derived analytic sequence.*

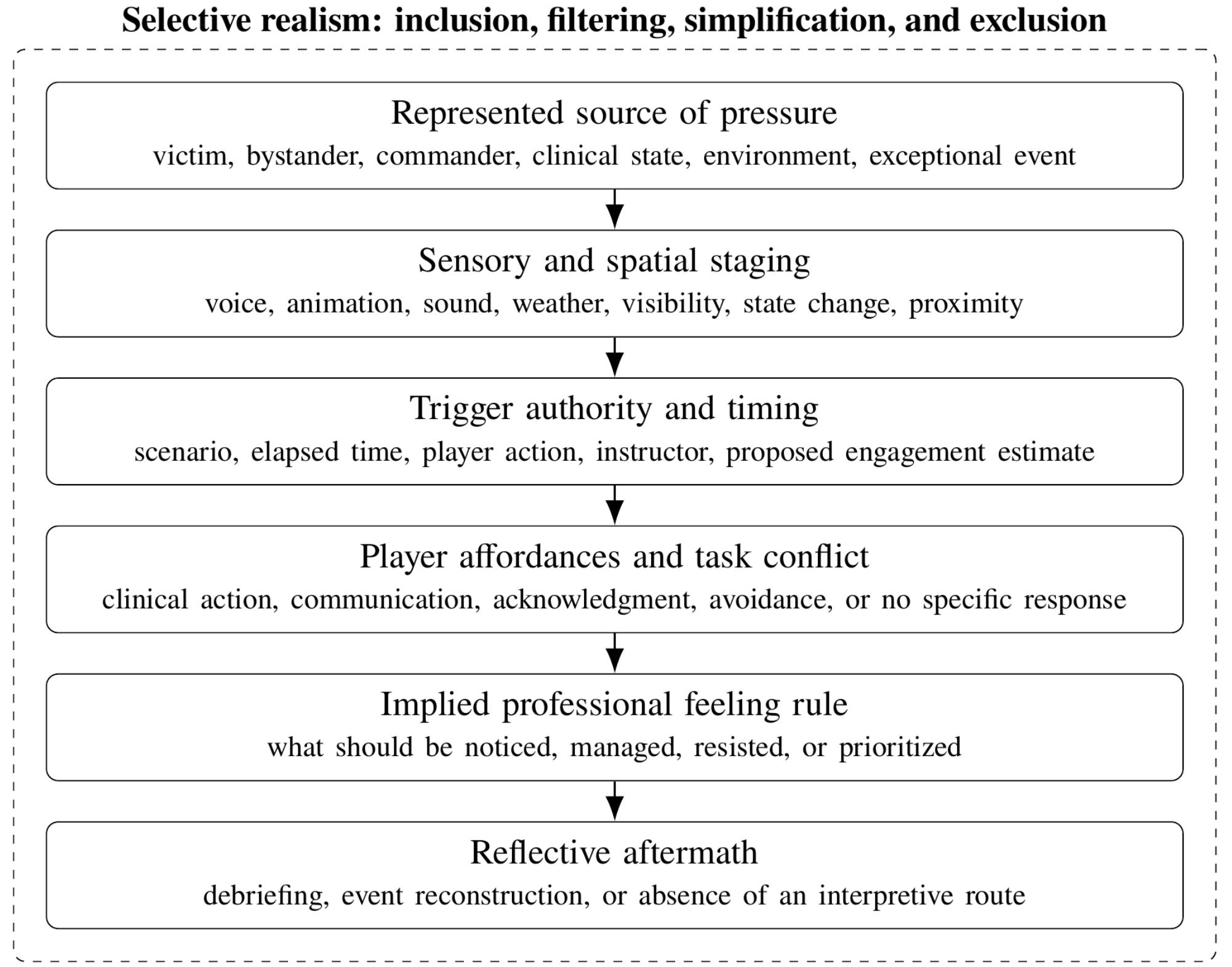

## Supplemental Appendix

*This appendix supplies blinded, derived descriptions needed to audit the article's selection and coding. The two proprietary source records are not reproduced in full. The appendix can be separated as online supplemental material at production.*

### Blinded Corpus Inventory

The focal corpus contains two contemporaneous Polish-language records: (a) a 34-page scenario-and-interaction specification; and (b) a seven-page system overview. The first documents the player role, VR interaction model, two airport settings, ten planned variants, character states, trigger rules, forty pattern specifications, and fuller minimum requirements for twenty prioritized patterns. The second describes the proposed bioadaptive architecture. No executable build, fixed walkthrough, interview, meeting record, participant data, or deployment observation is analyzed.

Figures 1–5 are project-archive development stills included only to illustrate visible scene composition. They are outside the two-record analytical corpus and do not evidence speech, sound, triggers, interaction, player attention, or user effects.

### Forty-Pattern Inventory

Status refers to the design specification, not runtime verification. 'Prioritized; minimum specification' identifies the twenty patterns selected for fuller requirements because the project expected clearer observable reactions. 'Inventory; not prioritized' identifies the remaining twenty documented candidates. Pattern descriptions are analytical paraphrases rather than quotations.

## Supplemental Table S1

*Derived forty-pattern inventory and design-stage status.*

| ID | Design-stage pattern | Status | Primary family or function and artifact-level note |
|---|---|---|---|
| P01 | Aggressive behavior by a casualty | Prioritized | Behavioral/interactional. Threatening speech and movement can be assigned to a conscious casualty; no differentiated de-escalation path is specified. |
| P02 | Aggressive behavior by a bystander | Prioritized | Behavioral/interactional. Makes interpersonal threat portable across bystanders and directs pressure toward persistence. |
| P03 | Commander observes and criticizes | Prioritized | Social scrutiny. Combines proximity, hierarchy, and public evaluation. |
| P04 | Screaming | Prioritized | Victim distress. Audible fear or pain can be activated on a conscious casualty as a portable claim on attention. |
| P05 | Crying | Prioritized | Victim distress. Audible grief can be assigned across casualties with limited biographical context. |
| P06 | Bystander comments | Prioritized | Social scrutiny. Bystanders hurry or criticize the rescuer, making task performance publicly accountable. |
| P07 | Hemorrhage | Prioritized | Clinical deterioration. A blood pool appears beneath a casualty; visual salience is clearer than a changing care pathway. |
| P08 | Phone recording | Prioritized | Social scrutiny. Phone models, flashes, and posting or streaming comments add reputational visibility. |
| P09 | Media presence | Prioritized | Social scrutiny. Reporters, camera operators, microphones, and questions create institutionalized observation. |
| P10 | Rain and storm | Prioritized | Environmental disruption. Dynamic weather, thunder, and lightning alter atmosphere without using a vulnerable person as carrier. |
| P11 | Fog | Prioritized | Environmental disruption. Reduced visibility may be task-relevant, but downstream procedural effects are thinly specified. |
| P12 | Presence of victims’ relatives | Prioritized | Victim distress. A nearby bystander is moved beside the user to plead for a relative, making proximity an intensity parameter. |
| P13 | Sudden death | Prioritized | Clinical deterioration. A casualty changes to a prone, nonbreathing, pulseless state; the documented change is irreversible. |
| P14 | Distant explosion | Prioritized | Exceptional spectacle. Deliberately conspicuous but placed far enough |

| | | | away not to alter local rescue action. |
|---|---|---|---|
| P15 | Giraffe | Prioritized | Exceptional spectacle. The records' improbable 'black swan' event exposes attentional capture as a purpose independent of task change. |
| P16 | Behavior framed in psychiatric terms | Prioritized | Behavioral/interactional. Hallucination-like behavior, disordered speech, and erratic movement form a portable mental-health-coded representation. |
| P17 | Panic in a casualty | Prioritized | Behavioral/interactional. Hyperventilation, fear, and refusal create a challenge without a dedicated care pathway. |
| P18 | Intrusive bystander | Prioritized | Social scrutiny. A bystander follows, asks personal questions, and offers irrelevant advice; boundary pressure exceeds boundary-management detail. |
| P19 | Early-pregnancy appeal | Prioritized | Victim distress. A conscious woman can demand priority for herself and an unborn child, making pregnancy a portable urgency marker. |
| P20 | Missing child | Prioritized | Victim distress. A casualty can plead for a missing child while holding a toy; the relation is vivid but no search protocol is specified. |
| P21 | Time pressure | Inventory; not prioritized | Procedural/temporal. Connects elapsed time to deterioration or hazard and locates pressure in the system rather than a person. |
| P22 | Call for help | Inventory; not prioritized | Victim distress/information. May direct attention toward a casualty or location and convey task-relevant information. |
| P23 | Open fracture or deep wound | Inventory; not prioritized | Clinical/visual severity. Uses visible injury and possible dialogue to increase bodily specificity and urgency. |
| P24 | Large number of casualties | Inventory; not prioritized | Resource/workload. Represents imbalance between demand and rescue capacity, foregrounding a systemic pressure. |
| P25 | Fog at night | Inventory; not prioritized | Environmental disruption. Layers visibility constraints and indicates combination among environmental conditions. |
| P26 | Snow | Inventory; not prioritized | Environmental disruption. Adds a weather-based visibility constraint. |
| P27 | Snow at night | Inventory; not prioritized | Environmental disruption. Combines weather and time-of-day constraints. |
| P28 | Rescuer's family member among casualties | Inventory; not prioritized | Relational/moral conflict. Could make impartiality and role conflict directly relevant. |
| P29 | Imprecise briefing | Inventory; not prioritized | Organizational/informational. Models moderate mismatch between institutional information and the scene. |
| P30 | Incorrect briefing | Inventory; not prioritized | Organizational/informational. Models severe mismatch, |

| | | | additional hazards, or major differences in casualty load. |
|---|---|---|---|
| P31 | Decapitation | Inventory; not prioritized | Bodily spectacle/clinical state. Highly graphic representation with unclear added action beyond classification. |
| P32 | Aircraft engine noise | Inventory; not prioritized | Environmental disruption. Can interfere with hearing and communication without assigning distress to a character. |
| P33 | Casualty in stupor | Inventory; not prioritized | Behavioral/clinical state. A minimally responsive casualty creates diagnostic ambiguity rather than overt volatility. |
| P34 | Crying bystander | Inventory; not prioritized | Victim distress/social environment. Extends grief beyond casualties and could support communication if answerable. |
| P35 | Fainting | Inventory; not prioritized | Clinical deterioration. A casualty or bystander collapses, potentially adding a new assessment demand. |
| P36 | Search for a child | Inventory; not prioritized | Relational claim/task implication. Unlike the prioritized appeal, this entry points toward locating a child and therefore possible action. |
| P37 | Unexpected health decline | Inventory; not prioritized | Clinical deterioration. Changes an already triaged casualty and makes reassessment central. |
| P38 | Fuel leak | Inventory; not prioritized | Environmental/dynamic hazard. Adds a material hazard with implications for scene safety and prioritization. |
| P39 | Threat to the rescuer's life | Inventory; not prioritized | Occupational safety. Makes self-protection and dynamic risk assessment explicit. |
| P40 | Amputation | Inventory; not prioritized | Bodily severity/clinical state. Represents severe injury through models, detached limbs, and possible dialogue. |

## Coding Guide

### Supplemental Table S2

*Operational fields used in the artifact analysis.*

| **Field** | **Operational definition** |
|---|---|
| Selection status | Whether the specification placed a pattern in the prioritized twenty or wider inventory; this does not infer implementation. |
| Source or carrier | The represented person, institution, body state, environment, or exceptional event that makes a claim on attention. |
| Sensory modality | Dialogue, nonverbal animation, sound, visual effect, weather, visibility, object appearance, or character-state change. |
| Spatial rule | Fixed location, nearest suitable character, movement near the user, ambient condition, or distant spectacle. |
| Trigger authority | Scenario start, elapsed time, player action, researcher/instructor decision, or proposed engagement estimate. |
| Temporal form | Onset, persistence, repetition, escalation, interruption, layering, or irreversible state change. |
| Task relationship | Direct clinical relevance, indirect professional relevance, material constraint, distraction, spectacle, or ambiguity. |
| Available response | Clinical procedure, generic or targeted communication, acknowledgment, avoidance, justified deferral, or no specific action. |
| Implied professional demand | The competence or orientation presupposed by the relation among pressure, task, and response. |
| Implied feeling rule | The artifact-level norm concerning what should be noticed, displayed, managed, resisted, or prioritized. |
| Selective-realism function | What is included, filtered, simplified, intensified, or excluded through representation and procedure. |
| In-play answerability | Whether a demand can be acknowledged, acted on, refused, delegated, recorded, or explicitly deferred during the scenario. |
| Post-play debriefability | Whether trigger, sequence, available options, rationale, and excluded actions can be reconstructed afterward. |
| Representational risk | Whether identity, vulnerability, or distress becomes interchangeable, stigmatizing, spectacular, or detached from care. |
| Confidence note | Whether an entry is explicit in a dated record, a close paraphrase, a translation judgment, or an interpretive inference. |

## Reflexive Audit Notes

*Scope.* The analysis is artifact-level. It excludes participant evidence and avoids claims that the design induced affect, improved performance, or validated learning.

*Specification.* Minimum requirements are reported as specifications. With no executable build or fixed walkthrough in the corpus, runtime appearance, sound, sequencing, and technical success are not inferred.

*Selection.* Anticipated observability is an explicit production criterion. The paper analyzes its distributional consequences while retaining feasibility, cost, sequencing, reuse, and evaluation convenience as alternatives.

*Families.* The six families are interpretive and exhaustive only for the prioritized twenty. They are not a psychological scale, designer-authored classification, or general taxonomy.

*Negative cases.* Weather, fog, explosion, and giraffe prevent reduction of pressure to victim distress and distinguish material constraints, atmosphere, and task-extraneous spectacle.

*Absence claims.* Teamwork, organizational constraints, recovery, and debriefing are described as comparatively weakly specified in the focal corpus, not absent from the wider project or real practice.

*Terminology.* Source language is paraphrased and contextualized. Analysis addresses the relation among framing, animation, reassignability, and response rather than treating relabeling as sufficient.

*Positionality.* Insider roles are withheld for double-anonymous review and must be restored in non-anonymized declarations. Dated records, stable IDs, negative cases, and logged inference levels constrain retrospective coherence.